\documentclass[a4paper,11pt]{article}
\usepackage{jinstpub}

\usepackage{amsmath,amsfonts}
\usepackage{algorithmic}
\usepackage{algorithm}
\usepackage{array}
\usepackage[caption=false,font=normalsize,labelfont=sf,textfont=sf]{subfig}
\usepackage{textcomp}
\usepackage{stfloats}
\usepackage{url}
\usepackage{verbatim}
\usepackage{graphicx}
\usepackage{orcidlink}
\usepackage{xcolor}
\usepackage{siunitx}
\usepackage{multirow}
\usepackage{makecell}
\usepackage{booktabs}
\usepackage{hyperref}
\usepackage{textalpha} 

\title{On-chip Bragg peak extraction from MHz frame rate X-ray detectors using a cellular automaton architecture}

\author[a,b,1]{S. Fowler,\note{Corresponding author.}}
\author[b]{S. Strempfer,}
\author[b]{D. Beniwal,}
\author[b,c]{M. Hammer,}
\author[b,c]{H. Shi,}
\author[b]{S. Gnanasekaran,}
\author[b,d]{N. Contini,}
\author[b]{T. Guruswamy,}
\author[c]{Y. Chen,}
\author[e]{L. Rota,}
\author[e]{D. Doering,}
\author[e]{A. Dragone,}
\author[b]{T. Zhou,}
\author[b]{M. Cherukara,}
\author[b]{K. Yoshii,}
\author[b,c]{and A. Miceli}

\affiliation[a]{University of Louisiana at Lafayette,\\ Lafayette, LA, U.S.A.}
\affiliation[b]{Argonne National Laboratory,\\ Lemont, IL, U.S.A.}
\affiliation[c]{University of Chicago,\\ Chicago, IL, U.S.A}
\affiliation[d]{Ohio State University,\\ Columbus, OH, U.S.A.}
\affiliation[e]{SLAC National Accelerator Laboratory,\\ Menlo Park, CA, U.S.A.}

\emailAdd{samantha.fowler1@louisiana.edu}

\abstract{
High-frame rate pixel detectors can produce data volumes that exceed available off-chip bandwidth, yet in many applications only a sparse subset of each frame carries relevant information. Spatially localized events, including diffraction peaks in crystallography, particle hits in tracking detectors, fluorescence spots in biological imaging, and other applications, all require that clusters of above-threshold pixels be identified and extracted from an otherwise featureless background. Conventionally this is performed in software algorithms such as connected-component labeling on full frames, but at \unit{\MHz} frame rates the resulting throughput becomes prohibitive. We present a lightweight hardware architecture that performs peak localization and patch extraction directly in the sensor silicon, transmitting only small pixel patches rather than complete frames. FPGA-based testing on an AMD Alveo V80 validated the synthesizability and timing closure of the peak-finding module in real hardware. The design replaces global connected-component labeling with a cellular automaton that uses purely local, fixed-iteration neighborhood operations, eliminating the label storage and equivalence-resolution logic that make conventional approaches impractical on-chip. We validate the architecture on X-ray Bragg peak detection for far-field high-energy diffraction microscopy and show that every peak found by a software reference is recovered, with equivalent downstream reconstructions. The architecture sustains several-hundred-\unit{\kHz} frame rates in \qty{130}{\nm} CMOS and exceeds \qty{1}{\MHz} in \qty{28}{\nm}.}

\begin{document}
\maketitle
\flushbottom

\section{Introduction}\label{sec:intro}
Modern imaging sensors in scientific, industrial, and defense applications are advancing rapidly in both pixel count and frame rate. Detectors operating at hundreds of kilohertz to megahertz frame rates are now deployed or planned across a range of domains, including X-ray and laser light sources, charged-particle tracking, time-resolved fluorescence microscopy, and high-speed machine vision. A recurring challenge across all of these domains is that the raw data rate far exceeds what can be practically transmitted off-chip, stored, or processed in real time by downstream computing infrastructure.

Fortunately, in many of these applications the information of interest is sparse: each frame contains a small number of localized, high-intensity events—bright spots, particle clusters, transient flashes—set against a large, relatively featureless background. If these events can be identified and extracted at the point of acquisition, only compact pixel patches surrounding each event need leave the sensor, achieving an order-of-magnitude or greater reduction in output bandwidth with no loss of scientifically relevant content. This amounts to a data reduction step executed in the sensor silicon itself, converting a bandwidth-limited system into one that scales with event rate rather than pixel count.

The canonical algorithmic tool for this task is connected-component labeling (CCL), a graph-theoretic method that groups contiguous above-threshold pixels into labeled regions \cite{Bolelli_OpenCV_CCA,ConnectedCompsAlgos,SparseCCL,GPUBased}. CCL is  widely implemented in software, where it underpins peak-finding routines in X-ray crystallography, blob detection in computer vision, and cluster finding in particle physics. Implementing CCL directly in hardware, however, presents significant difficulties. The algorithm requires per-pixel label storage, a global equivalence-resolution pass, and, in its most common two-pass form, random access to the full frame. These requirements translate into large SRAM footprints, complex control logic, and multi-pass data access patterns that are poorly matched to the streaming, area-constrained environment of a sensor ASIC. At MHz frame rates the processing of one frame must complete before the next arrives, a timing constraint that conventional CCL architectures struggle to meet.

In this work we propose an alternative architecture based on a cellular automaton (CA) approach. A CA based approach replaces global labeling with purely local neighborhood update rules applied over a fixed, small number of iterations. Connected regions of active pixels contract toward their spatial centers through these local interactions, ultimately converging to single representative pixels whose coordinates identify the events \cite{Palash_CA, CA_Survey_Jarkko, kari2022cellular, 2D_CA_Packard}. Because the CA operates with nearest-neighbor communication only, requires no label memory, and completes in a deterministic number of cycles, it maps naturally onto a compact, fully pipelined hardware datapath with fixed latency and predictable throughput. While CA based approaches have been used within FPGA designs for high energy physics particle tracking \cite{BAKHTERI20201999, 1303324, 1420908, DANTONE1999127}, this system is designed for ASIC and is tileable to allow for easy scalability. 

Although we use crystallographic Bragg peak detection as our primary validation vehicle, the architecture is agnostic to the physical origin of the events. Any imaging modality in which the signal of interest manifests as spatially localized clusters of elevated pixel values (e.g., particle tracking, fluorescence spot detection, transient event capture, or anomaly detection in industrial inspection) can benefit from the same on-chip data reduction strategy.

The principal contributions of this work are:
\begin{enumerate}
    \item A cellular-automaton formulation for peak localization that
          replaces global connected-component labeling with purely local,
          fixed-iteration neighborhood rules, eliminating per-pixel label
          storage and equivalence-resolution logic.
    \item A complete on-chip architecture---comprising the CA peak finder,
          an SRAM-based pixel pipeline, and a priority-merge output
          stage---that performs patch extraction at the point of acquisition,
          reducing output bandwidth by transmitting only small pixel patches
          rather than complete frames.
    \item ASIC synthesis results in both \SI{28}{\nano\meter} and
          \SI{130}{\nano\meter} CMOS demonstrating that the architecture
          sustains frame rates exceeding \SI{1}{\mega\hertz} and
          \SI{535}{\kilo\hertz}, respectively, within practical area and
          power budgets.
    \item An FPGA implementation on the AMD/Xilinx Versal platform
          providing a reconfigurable prototyping path.
    \item End-to-end validation on experimental far-field high-energy
          diffraction microscopy (FF-HEDM) data, showing that every Bragg
          peak identified by a conventional software connected-component
          pipeline is recovered by the CA architecture, and that downstream
          grain reconstructions via MIDAS are equivalent.
\end{enumerate}

The remainder of this paper is organized as follows. Section~\ref{sec:design} describes the design of the peak localization using a cellular automaton approach. Section~\ref{sec:asic_implementation} describe the hardware implementation . Section~\ref{sec:verifcation} describe the ASIC verification and FPGA-based testing validation and timing closure. Section~\ref{sec:evaluation} presents evaluation on a scientific workflow, and Section~\ref{sec:conclusion} concludes with a discussion of future directions.

\section{Design}\label{sec:design}
\begin{figure}[!htbp]
    \centering
    \includegraphics[width=0.5\textwidth]{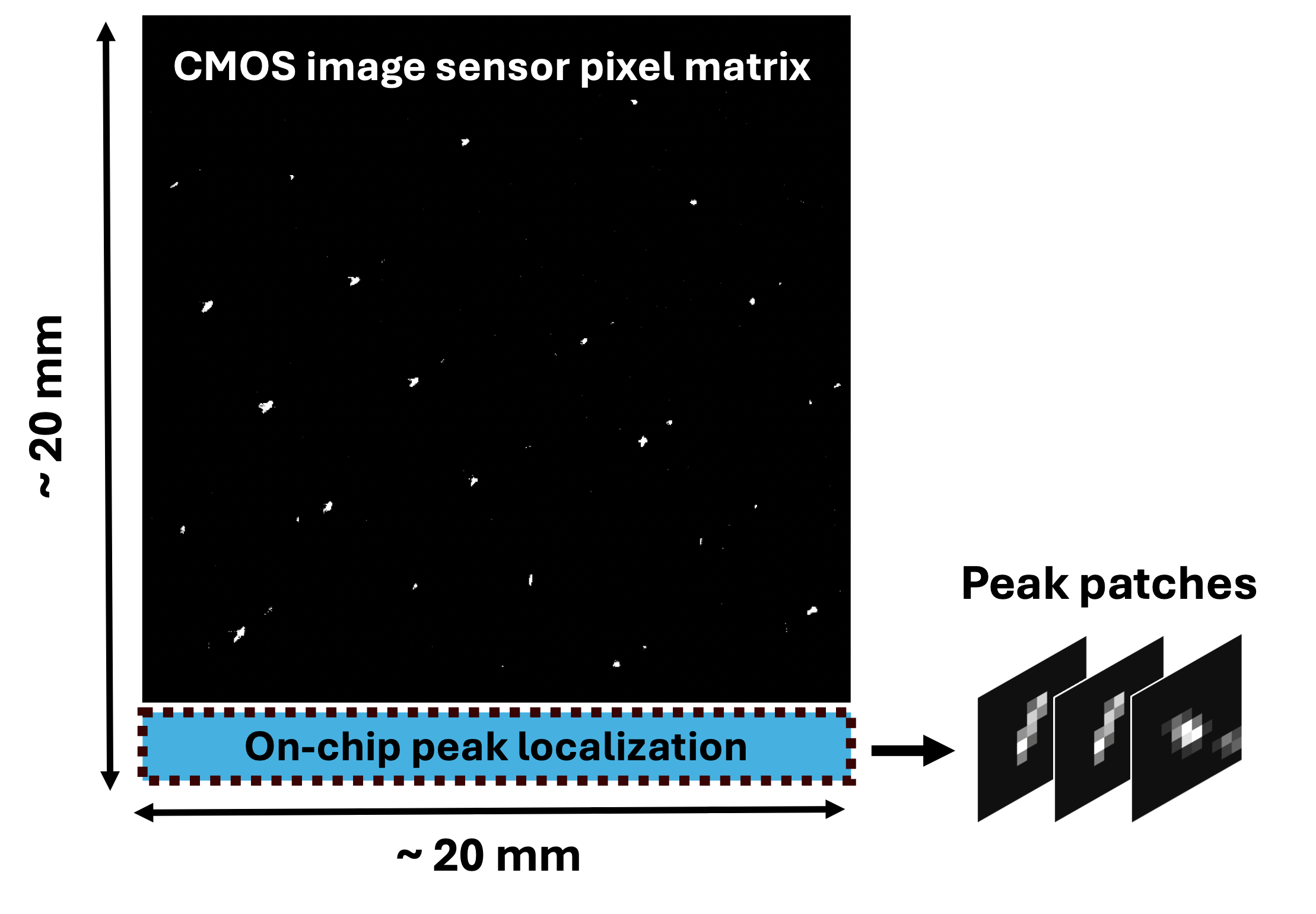} 
    \caption{Physical layout of the peak localization architecture integrated as a peripheral balcony at the edge of the sensor pixel matrix.}
    \label{fig:balcony}
\end{figure}

Co-locating peak localization with the sensor pixel matrix minimizes data
movement by performing one of the earliest data-reduction steps at the
point of acquisition. The architecture is designed to operate at the edge
of the pixel matrix, either integrated directly on the sensor die as a
peripheral \emph{balcony} or placed on a companion chiplet connected via
high-speed interconnects. In this paper we focus on the former
configuration, in which the processing logic resides on the sensor die
itself (Figure~\ref{fig:balcony}).

\begin{figure}[!htbp]
    \centering
    \includegraphics[width=0.75\textwidth]{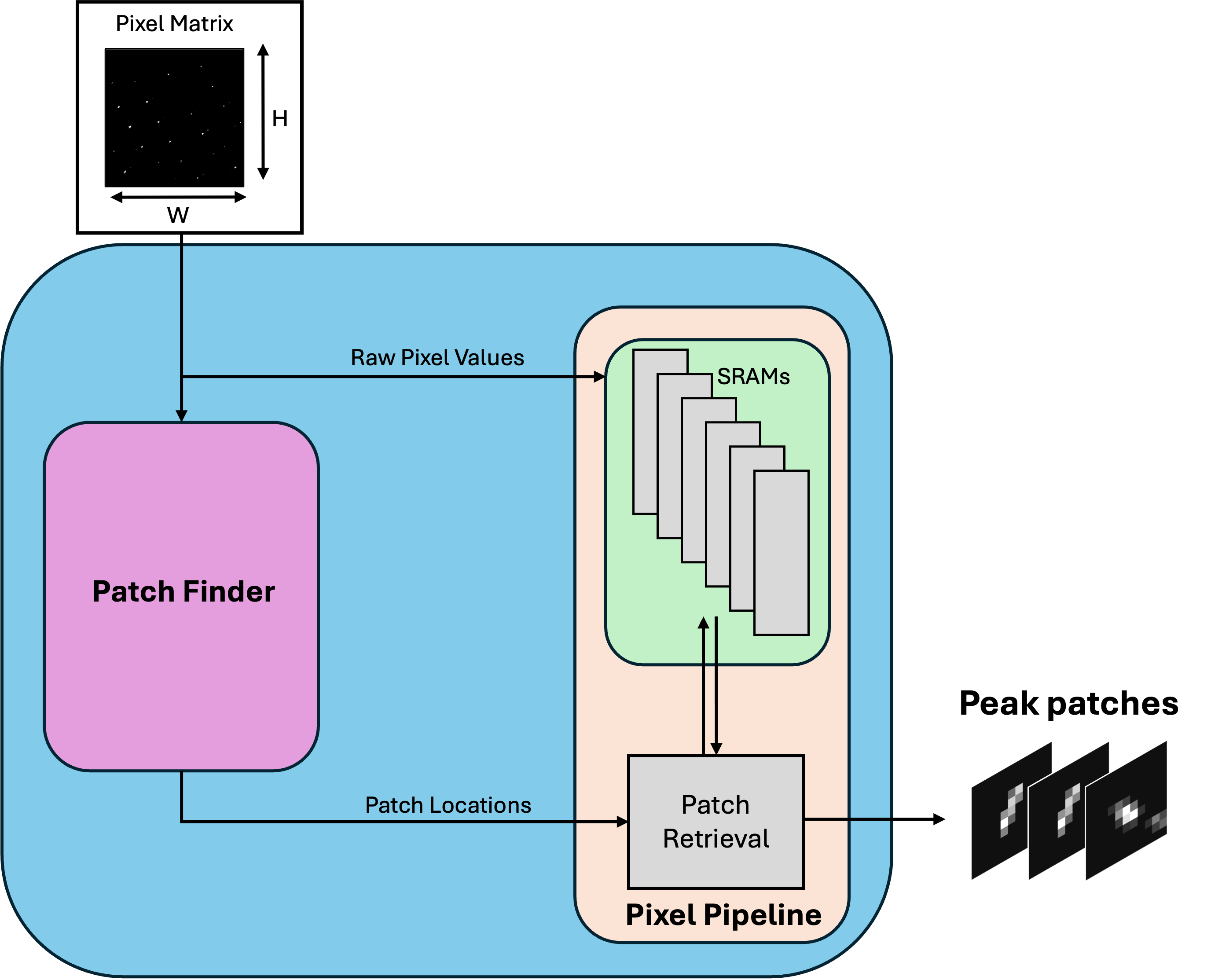} 
    \caption{Block diagram for overall peak location system, including both the pixel pipeline and the patch finder}
    \label{fig:overall-diagram}
\end{figure}

Pixel data is streamed row-wise from the matrix into the balcony, where
the patch-finder algorithm processes each frame before the next arrives.
As illustrated in Figure~\ref{fig:overall-diagram}, the architecture is composed of two primary modules: the \emph{patch finder} and the \emph{pixel
pipeline}. The patch finder operates directly on the incoming pixel stream
to identify peak locations in real time. Rather than storing full frames,
it analyzes the data as it arrives and outputs the coordinates of detected
peaks. In parallel, the pixel pipeline stores the raw pixel values across
multiple frames using SRAM buffers. Once peak locations are available, the
patch-retrieval module uses these coordinates to extract the corresponding
pixel regions from memory and stream them to the output.

\subsection{Cellular-Automaton--Based Peak Finding}
\label{sec:ca-algorithm}

The cellular automaton (CA) operates on a binary image obtained by
thresholding the raw detector frame. Each pixel becomes a \emph{cell}
whose state evolves synchronously over a fixed number of discrete
iterations~$K$. At every iteration each cell inspects only its \numproduct{3 x 3}
Moore neighborhood (i.e., the eight immediately surrounding cells) and updates
its own state according to a small set of local rules. No global
communication, label storage, or equivalence tables are required at any
point.

The key insight is that the combination of rules causes every connected
region of active pixels to \emph{erode inward} from its boundary while
simultaneously \emph{preserving} interior pixels that lie along medial
axes. After at most $K$ iterations each connected component has contracted
to a single surviving pixel located at or near the spatial center of the
original region. The coordinates of that surviving pixel are reported as
the peak location.

\begin{figure}[!htbp]
    \centering
    \includegraphics[width=0.75\textwidth]{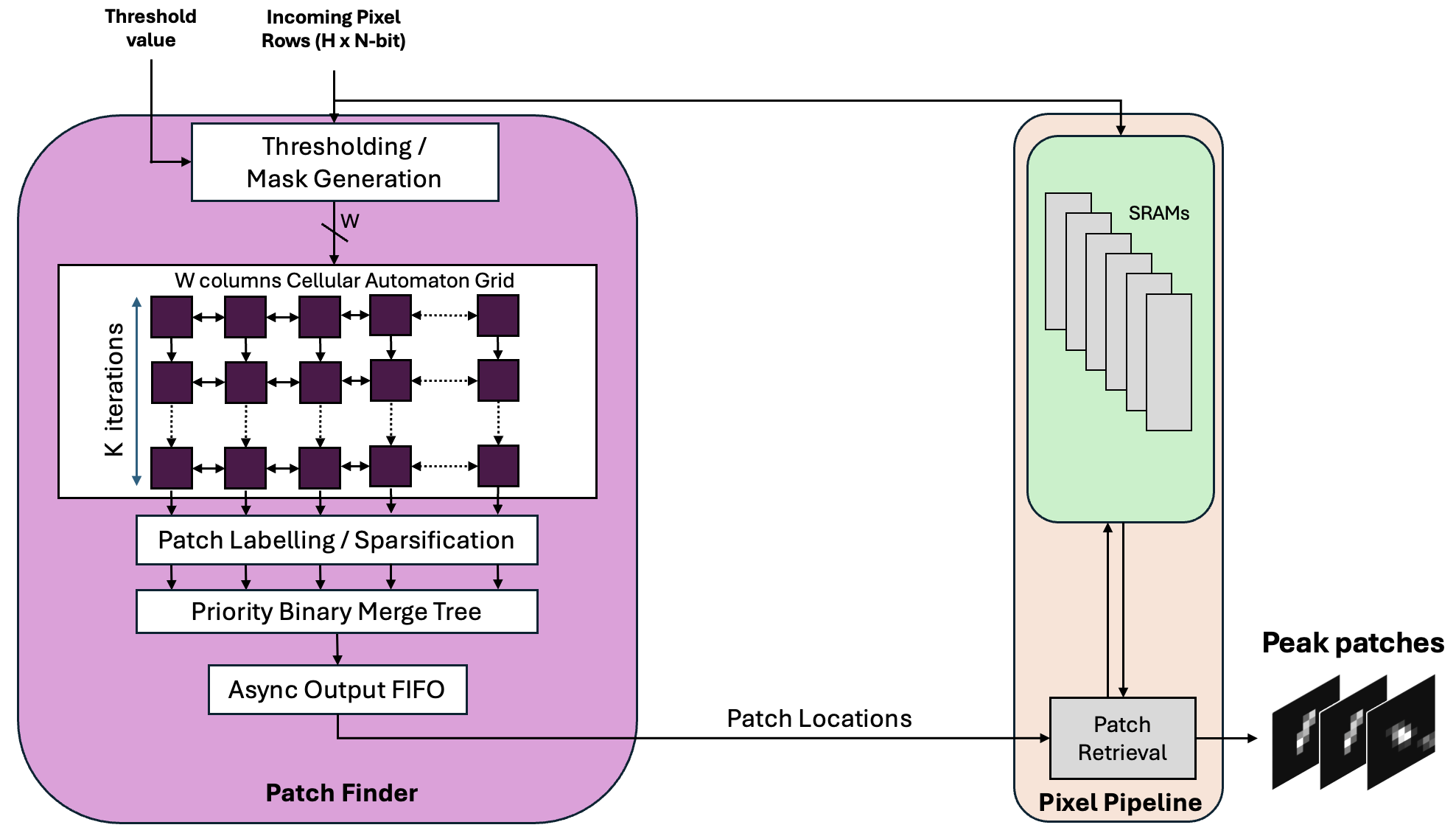} 
    \caption{Block diagram expanding upon the cellular automaton patch finder section of the system}
    \label{fig:CA-diagram}
\end{figure}

\subsubsection{Cell State}
\label{sec:cell-state}

Each cell carries a small state vector rather than a single bit.
Table~\ref{tab:cell-state} lists the fields. At the start of each frame, \texttt{value} is loaded from the thresholded pixel, \texttt{prevValue} is cleared, \texttt{isFirstRow} is asserted for the first row in each frame, and \texttt{iterations} is zero.

\begin{table}[!htbp]
\centering
\caption{Cell state vector}
\label{tab:cell-state}
\begin{tabular}{lp{1.2cm}p{10cm}}
\toprule
Field & Width & Description \\
\midrule
\texttt{value} & 1\,bit &
    Current binary state (active / inactive). \\[3pt]
\texttt{prevValue} & 1\,bit &
State from the previous iteration; used by the resurrection
rule. \\[3pt]
\texttt{isFirstRow} & 1\,bit &
Asserted for cells in the first row of each frame; enforces boundary
conditions so that the CA treats frame edges as surrounded by
inactive cells. \\[3pt]
\texttt{iterations} & 4\,bits\textsuperscript{*} &
Records the iteration at which the cell first became a
\emph{lone survivor}. Provides a coarse patch-size estimate;
optional in hardware. \\
\bottomrule
\multicolumn{3}{@{}l}{\footnotesize\textsuperscript{*}%
Width is $\lceil\log_2(K{+}1)\rceil$; 4\,bits for $K{=}11$.}
\end{tabular}
\end{table}

\subsubsection{Neighborhood Predicates}
\label{sec:predicates}

All update rules are expressed in terms of three Boolean predicates
computed from the \numproduct{3 x 3} Moore neighborhood. Each predicate maps to a
small combinational circuit in hardware.

\paragraph{Middle-of-a-line}
A cell satisfies this predicate if there exists \emph{at least one} axis
through the cell along which both opposing neighbors are active. Four axes
are tested: horizontal, vertical, and the two diagonals. 
If a cell has active neighbors on both sides along any of
these four directions, it is considered to lie on the interior skeleton of
a connected region and should survive the current erosion step. Boundary
pixels that protrude (i.e., having a neighbor on only one side of every
axis) do not satisfy the predicate and are eroded away. Notably this rule does \emph{not} check the center pixel itself, meaning dead pixels can be revived if surrounded by active cells. In hardware, this
is four two-input AND gates (one per axis) feeding a single OR gate.

\paragraph{Completely-alone}
A cell is completely alone if \emph{none} of its eight Moore neighbors is
active. This predicate identifies cells that have either always been
isolated single-pixel events, or have been reduced to a single surviving
pixel by the erosion process (i.e., converged patch centers). In hardware,
this is an eight-input NOR gate.

\paragraph{Between-two-patches}
This predicate detects \emph{inactive} cells that sit in the gap between
two or more distinct active regions. It prevents the middle-of-a-line rule
from erroneously bridging separate patches. The eight perimeter neighbors are read in clockwise order starting from the top-left corner, 
and their active/inactive states are treated as a
circular binary string. The number of $0{\to}1$ and $1{\to}0$
\emph{transitions} in this string is counted. Two transitions correspond
to a single contiguous run of active neighbors (one connected region);
more than two transitions imply that multiple distinct regions surround
the cell. The predicate fires when the center cell is inactive and the
transition count exceeds two. In hardware this requires an 8-bit XOR
chain of adjacent perimeter bits followed by a population count comparator
(approximately 30~gates).

\subsubsection{Update Rules}
\label{sec:rules}

At each iteration, every cell computes its next state as the result of
three rules applied in parallel. If \emph{any} rule produces an active
result, the cell is active in the next iteration.

\paragraph{$\boldsymbol{R}_\textbf{erode}$ (primary rule)}\label{R-erode}
A cell survives if it satisfies \texttt{middle\_of\_a\_line} \emph{and}
does \emph{not} satisfy \texttt{between\_two\_patches}. This is the core component of the algorithm: 
it preserves the medial axis of each connected
region while peeling away boundary pixels one layer per iteration. The
\texttt{between\_two\_patches} guard is crucial; without it, a dead
pixel flanked by two independent patches would be spuriously activated,
merging the patches into one.

\paragraph{$\boldsymbol{R}_\textbf{keep}$ (lone-survivor preservation)}\label{R-keep}
An active cell that has no remaining active neighbors (i.e., it is
\texttt{completely\_alone}) is already a converged center. This rule
prevents it from being eroded away in subsequent iterations. Once a patch
has shrunk to a single pixel, that pixel is ``locked in'' and persists
through all remaining iterations.

\paragraph{$\boldsymbol{R}_{\textbf{resurrect}}$ (simultaneous-death recovery)}\label{R-resurrect}
This rule handles a subtle edge case that arises with even-sized patches.
When a connected region has an even diameter (e.g., a \numproduct{2 x 2} square),
all remaining pixels may satisfy the erosion rule simultaneously and die
together in the same iteration, leaving no survivor. 
The resurrection rule detects this event by checking four conditions
simultaneously: (1)~the cell was active in the \emph{previous} iteration,
(2)~it is now dead, (3)~it is completely alone (all neighbors also died),
and (4)~none of the four neighbors in the \emph{lower-right corner}
(bottom-left, bottom-center, bottom-right, and center-right) were previously
active. The fourth condition is an asymmetric tie-breaker: because only the
bottom-right cell of any compact dying group has all four bottom-right
neighbors inactive, exactly one cell per vanished group is resurrected.
This deterministic tie-breaking avoids duplicate detections without
requiring any global coordination. The ASIC implementation has this rule vertically flipped since frames are processed bottom-to-top.

\subsubsection{Worked Example}
\label{sec:worked-example}
\begin{figure}[!htbp]
    \centering
    \includegraphics[width=\textwidth]{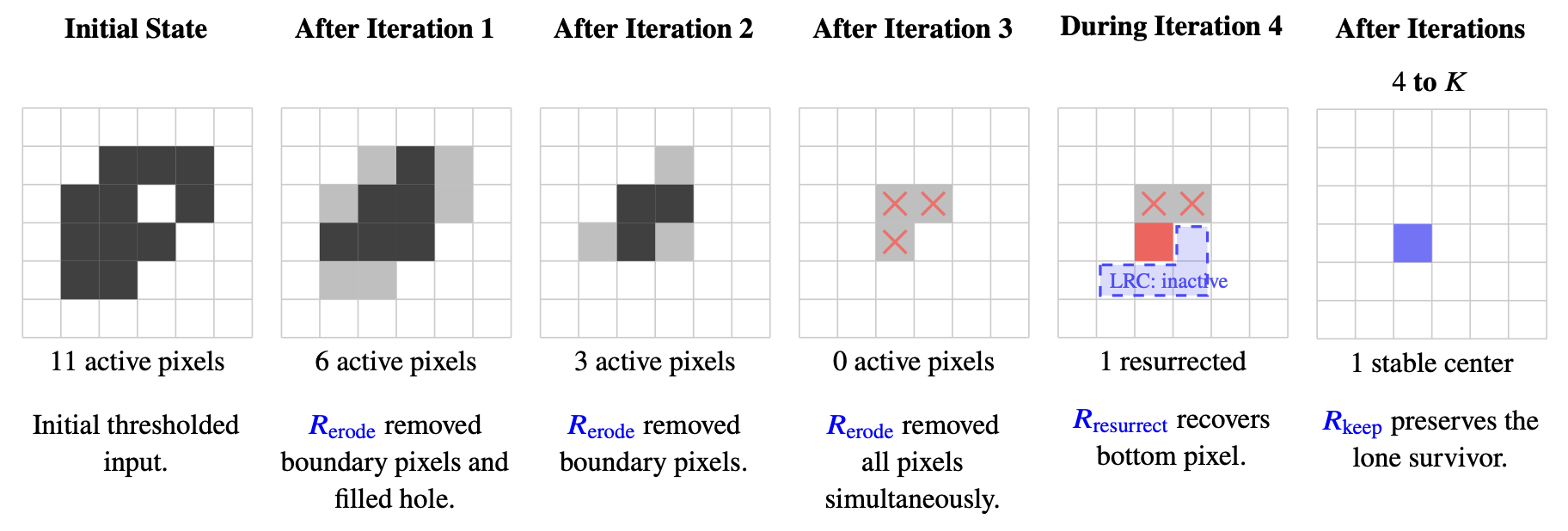} 
    \caption{Evolution of a \numproduct{4 x 4} active region through the CA rules.
Light gray squares show pixels that were active in the previous iteration
but have been eroded. Red crosses mark simultaneous deaths. The dashed
blue box in iteration~4 highlights the lower-right
corner (LRC) neighbors checked by the resurrection rule for the bottom pixel; because all four
are inactive, the bottom pixel is the unique cell that satisfies the
resurrection condition. This rule is also checked for the other two cells but is not satisfied (not shown). From iteration~4 onward, the surviving pixel is
completely alone and is held stable by $R_{\text{keep}}$. Its coordinates
are reported as the peak center.}
    \label{fig:worked-example}
\end{figure}

Figure~\ref{fig:worked-example} traces the evolution of a \numproduct{4 x 4}
patch through the CA rules, illustrating all three rules in action. The example illustrates several important properties:
\begin{itemize}
\item \textbf{Erosion rate.} Approximately half a layer of boundary pixels is removed
per iteration, so a \numproduct{4 x 4} patch reduces to \numproduct{3 x 3}, then \numproduct{2 x 2} in consecutive iterations.
\item \textbf{Even-size handling.} The \numproduct{2 x 2} residual has no
unique geometric center; all pixels die simultaneously. The
resurrection rule recovers exactly one pixel using the asymmetric
LRC tie-breaker.
\item \textbf{Deterministic convergence.} The patch converges in
4~iterations regardless of its position in the frame. For an
odd-sized patch such as \numproduct{3 x 3}, no resurrection is needed:
the single center pixel survives erosion directly and is then
preserved by $R_{\text{keep}}$.
\item \textbf{Center accuracy.} The reported center $(2,2)$ is
within one pixel of the geometric center of mass $\mathtt{\sim}(2.3, 2.6)$,
consistent with the sub-pixel accuracy observed in the validation
experiments (Section~\ref{subsec:ASICVer}).
\end{itemize}

\subsubsection{Minimum Patch Size Filtering}
\label{sec:min-patch}

Not all three rules are active in every iteration. A configurable
\texttt{min\_patch\_size} parameter controls early-iteration behavior to
suppress noise-induced small detections without post-processing.
Table~\ref{tab:min-patch} summarizes the activation schedule.
The logic is straightforward: disabling $R_{\text{keep}}$ at iteration~0
means that a single isolated active pixel is not preserved and dies,
filtering out 1-pixel noise hits. Disabling $R_{\text{resurrect}}$ at
iteration~1 means that a \numproduct{2 x 2} cluster---which would die
simultaneously and then resurrect---is not recovered, filtering out
2-pixel clusters. A \texttt{min\_patch\_size} of \numproduct{3x3} pixels cannot reliably be enforced by disabling rules and is therefore not supported. In hardware, \texttt{min\_patch\_size} can optionally be
exposed as a run-time configurable input.

\begin{table}[!htbp]
\centering
\caption{Rule activation schedule as a function\\of the
\texttt{min\_patch\_size} parameter}
\label{tab:min-patch}
\begin{tabular}{cccc}
\toprule
\texttt{min\_patch\_size} &
Iter.\ 1 & Iter.\ 2 & Iter.\ $\geq 3$ \\
\midrule
1 (report all) &
all three rules & all & all \\
2 (suppress 1$\times$1-px) &
$R_{\text{erode}}$ only & all & all \\
3 (suppress $\leq$2$\times$2-px) &
$R_{\text{erode}}$ only &
$R_{\text{erode}}$, $R_{\text{keep}}$ &
all \\
\bottomrule
\end{tabular}
\end{table}

\subsubsection{Convergence and Iteration Count}
\label{sec:convergence}

The erosion process removes at most one layer of boundary pixels per iteration. A connected region with a bounding box of $w\times h$ pixels therefore requires at most $K = \lfloor (w + h)/2 \rfloor + 1$ iterations to contract to a single pixel.
In general, the number of iterations is related to the hamming distance from the center to the furthest pixel, with $R_{\text{resurrect}}$ adding an additional iteration when necessary. For the target application
(i.e., Bragg peaks up to \numproduct{10 x 10} pixels), $K=11$ suffices. Because $K$ is
fixed at design time, the CA completes in exactly $K$ clock cycles per row
regardless of input data, yielding deterministic latency and throughput. When a cell first becomes a lone survivor, the current iteration index is
recorded in the \texttt{iterations} field. This value serves as a coarse
proxy for patch size: larger patches converge later. The field is reported
alongside the peak coordinates.

\subsubsection{Final-Stage Cell}
\label{sec:final-stage}

The last iteration stage ($i = K$) uses a simplified rule that outputs
\emph{only} fully converged peaks: only $R_{\text{keep}}$ and
$R_{\text{resurrect}}$ are active. The erosion rule is omitted because any
pixel that still has active neighbors after $K$ iterations belongs to a
patch larger than the design maximum and should not generate a spurious
detection. This ensures that the output is a sparse binary mask in which
each remaining active pixel corresponds to exactly one detected peak.

\subsubsection{Hardware Mapping}
\label{sec:hw-mapping}

The CA maps directly to a systolic array of $W \times K$ identical cells
(one column per pixel, one row per iteration), as shown in
Figure~\ref{fig:CA-diagram}. Pixel data streams through the array one
detector row per clock cycle in a bottom-to-top order. Each cell contains
two registers holding the current and previous row states; together with
the incoming pixel from the upstream stage, these form the center column
of the \numproduct{3 x 3} neighborhood. To avoid long combinational logic chains, the output of each cell is also registered, meaning in total each cell needs registers to store three of the cell states shown in Table~\ref{tab:cell-state}. The left and right columns come from the
corresponding registers of the neighboring cells in the same iteration
stage.

This nearest-neighbor-only connectivity eliminates long wires and enables
straightforward place-and-route. All three predicates are purely
combinational functions of the \numproduct{3 x 3} neighborhood, so each cell is a
small combinational block followed by a single register stage. An enable
signal cascades through the iteration stages with one cycle of delay per
stage, providing automatic pipeline flushing at frame boundaries.

\subsubsection{Patch Sparsification and Priority Merge Tree}
\label{sec:merge-tree}

Following the CA evolution, the $W$-wide output row is scanned for
remaining active pixels. Each active pixel generates a patch record
containing its column index (center~$x$), the current row counter
(center~$y$), the frame number, and optionally the \texttt{iterations}
field as a size estimate.

Because multiple peaks may be detected in the same row, a binary merge
tree consolidates the $W$ single-pixel outputs into a single ordered
stream. The tree has $\lceil\log_2 W\rceil$ levels; at each level a
two-input arbiter selects between its left and right children. When
sorting by $y$-coordinate is enabled, the arbiter prioritizes the patch
with the smaller (frame number, center~$y$) tuple, ensuring that the
output stream is sorted in raster order. Each merge node
contains a small FIFO to absorb transient throughput mismatches. The root of the merge tree feeds an asynchronous FIFO that bridges the
detection clock domain to the downstream patch-retrieval clock domain,
decoupling the CA pipeline from memory-access timing.

\subsection{Pixel Pipeline}
\begin{figure}[!htbp]
    \centering
    \includegraphics[width=0.75\textwidth]{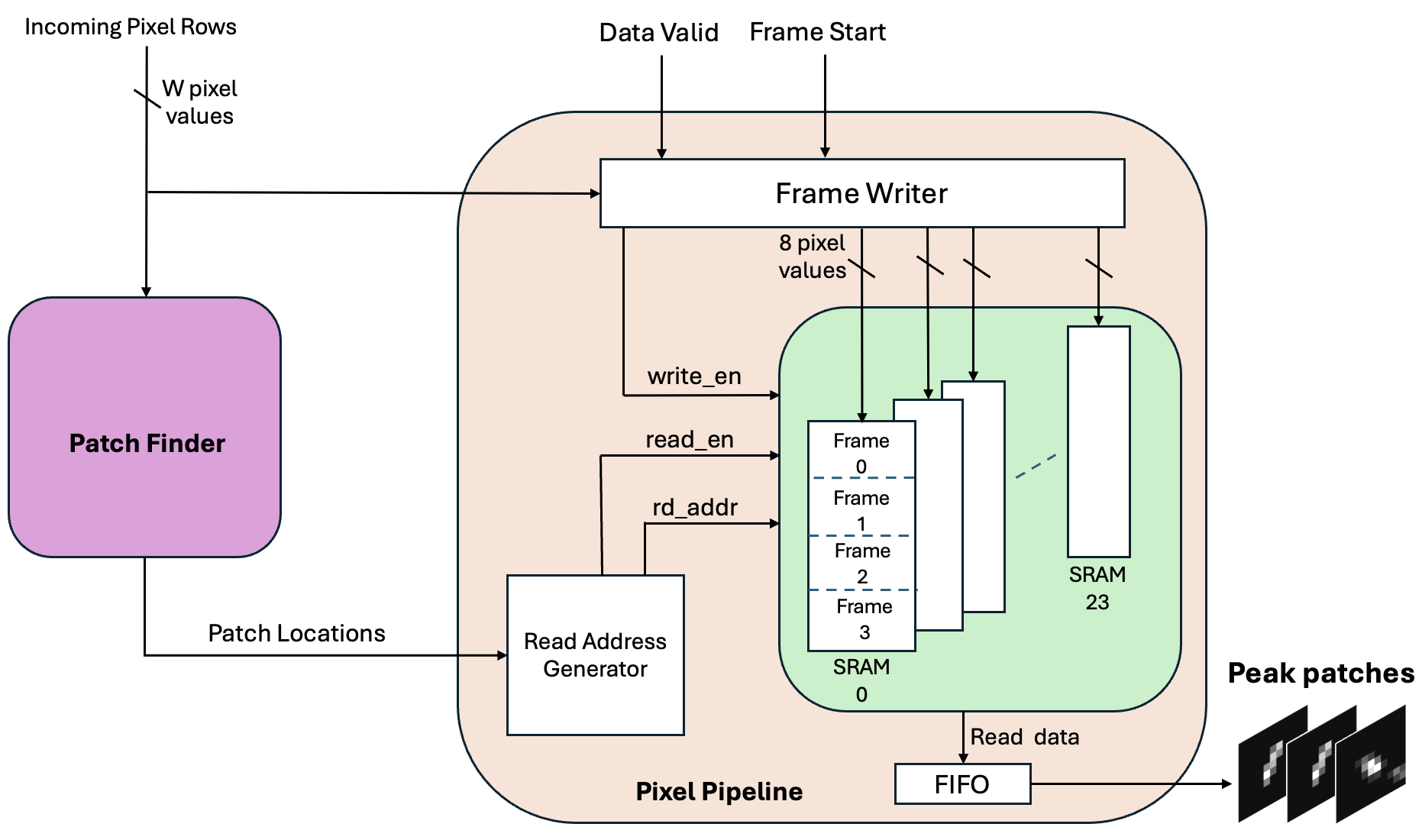} 
    \caption{Block diagram expanding upon the pixel pipeline section of the system}
    \label{fig:PP-diagram}
\end{figure}

The pixel pipeline architecture, shown in Figure~\ref{fig:PP-diagram}, is responsible for buffering incoming detector data and enabling random-access patch extraction. The design consists of SRAM banks, collectively capable of storing four complete frames. Each SRAM word contains eight 12-bit pixels, allowing an entire row of pixels to be written in parallel across all banks in a single cycle.
Incoming pixel rows are written into the SRAM array through the frame writer module, which operates at the detector input rate and manages frame indexing and address generation. In parallel, a read address generation module receives patch center coordinates and frame indices from the cellular automaton-based peak finder and computes the corresponding memory addresses required for patch extraction.
Due to the banked memory organization, each SRAM stores only a subset of pixels within a row. As a result, extracting a patch requires coordinated access across multiple SRAM banks. The read address generator determines the bank containing the center pixel and issues read requests to the appropriate neighboring banks to reconstruct the full spatial window. In order to avoid read conflicts in experiments when events are dense and localized, patches are extracted one at a time. Retrieved data is then assembled and streamed through a FIFO interface to produce the final patch outputs.

\section{ASIC implementation}\label{sec:asic_implementation}
As a realistic design target, we adopt the ePixUHR pixel matrix~\cite{ePixUHR}, which produces \numproduct{192 x 168} pixel frames. Each frame is streamed into the on-chip processing logic one 192-pixel row per cycle in a bottom-to-top order. As shown in Figure~\ref{fig:overall-diagram}, incoming pixels are routed simultaneously to the thresholding front-end of the patch finder and to the SRAM-based pixel pipeline, preserving the original pixel values for subsequent patch extraction. The pixel pipeline employs a banked memory architecture consisting of 24 SRAMs, each 629 words deep and 96 bits wide. Collectively, the array stores four complete detector frames, providing sufficient buffering for the CA-based patch finder to complete peak detection before any frame is overwritten. Once a peak is localized, its coordinates and frame index are used to extract a fixed-size \numproduct{11 x 11} pixel patch from the stored data. For this \numproduct{11 x 11} pixel patch size, the number of iterations performed by the CA algorithm is set to 11. This ensures that all peak events \numproduct{11 x 11} or smaller can be found. The system was also set to ensure that the smallest reported peak would have 3 pixels minimum in either the x or y direction, to reduce noise within the output.

The architecture uses two clock domains to decouple data ingestion from patch retrieval. The SRAM write interface and the patch finder operate at \qty{168}{\MHz}, matching the detector row rate and yielding a frame rate of \qty{1}{\MHz}. The patch retrieval module operates at \qty{504}{\MHz} ($3\times$ the write clock) to sustain the bandwidth required for random-access reads across multiple SRAM banks. An asynchronous FIFO bridges the two domains, buffering detected peak coordinates until the retrieval module can service them. This FIFO has a depth of 8, allowing for a back-up of 8 patches before dropping subsequent patches. The number of dropped patches is tracked and reported to the user. Upon receiving a peak's center coordinates and frame number from the FIFO, the retrieval module identifies the SRAM bank containing the center pixel and issues coordinated reads to the neighboring banks. The returned data are assembled and cropped to produce the \numproduct{11 x 11} output patch, centered on the detected peak. 

The design was synthesized in both a \qty{28}{\nm} and a \qty{130}{\nm} CMOS process. Tables~\ref{tab:asic-clock} and~\ref{tab:asic-area-power} summarize the timing, area, and power results. At \qty{28}{\nm} the design meets timing at the target \qty{168}{\MHz}\,/\,\qty{504}{\MHz} clocking, sustaining a \qty{1}{\MHz} frame rate and occupying only \qty{0.64}{\square\mm}---well within the \qty{40}{\square\mm} area budget available on the sensor die. At \qty{130}{\nm} the critical path limits the maximum clock to \qty{270}{\MHz}; because the read clock must remain $3\times$ the write clock, the effective write clock is \qty{90}{\MHz}, reducing the frame rate to approximately \qty{535}{\kHz}. The \qty{130}{\nm} implementation requires \qty{31.78}{\square\mm}, consuming most of the available die area. In both nodes SRAM dominates the total area (\qty{82.8}{\percent} at \qty{28}{\nm}, \qty{94.9}{\percent} at \qty{130}{\nm}), confirming that on-chip frame buffering rather than CA logic is the primary area cost. This imbalance will grow with pixel count or buffer depth, since the CA footprint remains essentially fixed. Process scaling also strongly affects power. The \qty{28}{\nm} implementation dissipates \qty{6.6}{\mW} total versus \qty{846.3}{\mW} at \qty{130}{\nm}. Leakage is higher at \qty{28}{\nm} (\qty{6.9}{\mW} vs.\ \qty{1.9}{\mW}) due to increased subthreshold conduction, but this is far outweighed by the reduction in dynamic power, making the \qty{28}{\nm} implementation the more suitable target for high-throughput detector systems.

\begin{table}[!htbp] \centering \caption{ASIC Timing Performance} \label{tab:asic-clock} \begin{tabular}{c S[table-format=+1.3] S[table-format=3] S[table-format=1.3]} \hline \multirow{2}*{Process Node} & {Worst Slack} & {Max Clock} & {Max Frame Rate} \\ & {(\unit{\ns})} & {(\unit{\MHz})} & {(\unit{\MHz})} \\ \hline \qty{28}{\nm} & {Met Timing} & 504 & 1.0\\ \qty{130}{\nm} & -1.703 & 270 & 0.535\\ \hline \end{tabular} 
\end{table} 

\begin{table}[!htbp] \centering \caption{ASIC Area and Power Results} \label{tab:asic-area-power} 
\resizebox{\columnwidth}{!}{ 
\begin{tabular}{c | S[table-format=2.2] S[table-format=1.2] S[table-format=2.2] S[table-format=2.1] | S[table-format=3.1] S[table-format=3.1] S[table-format=3.1] S[table-format=1.1]} \hline 
\multirow{3}*{Process Node} & \multicolumn{4}{c|}{Area} & \multicolumn{4}{c}{Power} \\ & {Total} & {Logic} & {SRAM} & {SRAM} & {Total} & {Logic} & {Periph.} & {Leakage} \\ & {(\unit{\square\mm})} & {(\unit{\square\mm})} & {(\unit{\square\mm})} & {(\unit{\percent})} & {(\unit{\mW})} & {(\unit{\mW})} & {(\unit{\mW})} & {(\unit{\mW})}\\ \hline 
\qty{28}{\nm} & 0.64 & 0.11 & 0.53 & 82.8 & 6.6 & 1.7 & 4.9 & 6.9 \\ \qty{130}{\nm} & 31.78 & 1.61 & 30.16 & 94.9 & 846.3 & 235.7 & 610.6 & 1.9 \\ \hline 
\end{tabular} 
} \end{table}

\section{Verification}\label{sec:verifcation}
This section will detail the verification methods used for the ASIC implementation which was verified both functionally and timing wise. We also present a FPGA implementation for integration validation.

\subsection{ASIC Verification}\label{subsec:ASICVer}
The ASIC implementation was evaluated using a cocotb based simulation consisting of 15 frames at \numproduct{192x168} pixels containing 10 peaks per frame. In order to accommodate these images, an SRAM size of 629 words by 96 bits was selected. Across all 150 generated peaks, the system successfully detected each peak and properly extracted them as \numproduct{11 x 11} patches. This evaluation verified that the multi frame SRAM buffering prevented frames from being overwritten before patch extraction completed, confirmed that all peaks were successfully localized, and demonstrated that the architecture could sustain the required throughput under continuous streaming operation.

To further verify the proposed architecture, a software model was developed to emulate the behavior of the hardware implementation. Experimental data from samples scanned at multiple pressures were used to assess performance as the peaks broaden and elongate at higher pressures \cite{peakBroadening}.

The peak centers identified by the proposed cellular automaton (CA)-based method were compared against those obtained from a reference software pipeline based on connected components labeling. A matching tolerance of 10 pixels was used to associate detections between the two methods. The results show that all peaks identified by the reference implementation are also detected by the CA-based approach, indicating strong agreement in core peak localization.

In addition, the CA based method produces a number of extra detections not present in the reference output. These additional peaks primarily arise from differences in filtering criteria between the two approaches. Specifically, the proposed implementation does not enforce constraints on peak size relative to the fixed \numproduct{11 x 11} patch window, and therefore retains detections corresponding to larger or partially overlapping structures that are discarded in the connected components-based method. As a result, the CA-based system favors higher sensitivity at the expense of increased redundant or oversized detections. Overall, these results demonstrate that the proposed approach achieves full coverage of relevant peak locations while maintaining robustness to variations in peak shape, making it well suited for hardware-based, streaming implementations.

\subsection{FPGA-based Testing and Validation}\label{subsec:FPGAVer}

The goal of this FPGA test was implementation and integration validation, not final FPGA performance optimization. While software simulation verifies cycle-level functional correctness under modeled conditions, this FPGA-based testing confirms that the design can be physically implemented, timing-closed, integrated with the host/PCIe/AXI control path, and executed repeatedly in a realistic hardware environment.

We verified the RTL implementation of the Cellular-Automaton–Based Peak Finding module, a logically complex part of the pixel pipeline, on an AMD Alveo V80 FPGA using Vivado 2025.1. The test framework was built using AVED, the AMD/Xilinx Alveo Versal Example Design framework.

In this setup, the peak finding RTL is integrated into an FPGA-side test harness and wrapped with an AXI dispatching module. Host software running on the CPU sends commands over PCIe, which arrive at the FPGA test module as AXI4 transactions. These commands write input data into the input FIFO, start data injection into the peak finding module, and read results from the output FIFO.

The RTL achieved timing closure at a 10 ns clock period. The reported critical path delay was 3.133 ns, consisting of 0.740 ns logic delay and 2.393 ns routing delay, indicating that the path was primarily routing-dominated. The peak finding module together with the test harness used 742 registers and 1,512 CLB LUTs, including 904 LUTs as logic and 608 LUTs as LUT-based memory, while using no BRAMs. This corresponds to well below 1\% of the available FPGA resources.

\section{Evaluation on a Scientific  Workflow -- X-ray diffraction microscopy}\label{sec:evaluation}
To validate the proposed architecture end-to-end, we evaluate it on a scientific workflow called far-field high-energy diffraction microscopy (FF-HEDM \cite{annurev-hedm, jsp-hedm,Lienert2011}) workflow. FF-HEDM is a non-destructive technique used at synchrotrons to characterize the internal microstructure of polycrystalline materials, such as metals and alloys. A high-energy X-ray beam illuminates a polycrystalline sample while the sample is incrementally rotated about an axis $\omega$. At each rotation angle, crystallographic planes within individual grains diffract the beam according to Bragg's law, producing bright spots (Bragg peaks) on a downstream area detector. By collecting diffraction frames across the full $\omega$ sweep, the position, orientation, and elastic strain state of every grain in the illuminated volume can be reconstructed.

We use the Microstructural Imaging using Diffraction Analysis Software (MIDAS) software \cite{sharma_marinerhemantmidas_2026} to analyze FF-HEDM data. Given a stack of diffraction frames, MIDAS performs the following steps: (i) connected-component segmentation to isolate Bragg peaks in each frame, (ii) 2-D multi pseudo-Voigt fitting of each peak in detector polar coordinates (R,$\eta$) to extract its centroid, integrated intensity, and shape, (iii) ring-by-ring indexing, in which the fitted spots are matched against a candidate orientation grid to assign each peak to a grain, and (iv) Levenberg--Marquardt refinement of the position, orientation, and elastic strain tensor for each grain. The final output is a grain map: a catalog of all grains in the illuminated volume with their crystallographic orientations and strain states.

MIDAS requires only the per-frame peak list from step (ii) as input for subsequent indexing and refinement stages. It is therefore agnostic to how the Bragg peaks were originally segmented. This means MIDAS can be driven equally well by patches extracted from a conventional software connected-component analysis on full detector frames or by the compact patches produced by the on-chip CA-based peak localization proposed here. We exploit this modularity to perform a direct comparison: we run two parallel MIDAS reconstructions on the same diffraction dataset, one using software-CCL patches from full frames and one using ASIC-CCL patches from the on-chip architecture, and compare the resulting grain maps. Any difference in reconstruction quality is therefore attributable solely to the upstream peak segmentation step.

\subsection{MIDAS Pipeline Construction}
\label{subsec:pipeline-construction}

To isolate the CCL stage as the only variable in the comparison, we construct two pipelines that share \emph{every} downstream component: the same intensity preprocessing, same pseudo-Voigt peak fitting routine, and the same MIDAS indexing and refinement back-end with the same parameter file. The two pipelines differ only in where the input pixels come from:

\begin{itemize}
    \item \textbf{Software-CCL pipeline.} The full detector frame is
          dark-subtracted, ring-masked, and intensity-thresholded.
          Pixels surviving these steps are passed to software CCL
          (\texttt{scipy.ndimage.label}), which emits an
          \numproduct{11 x 11} window centered on each connected
          region. These windows are fit with a pseudo-Voigt routine to produce a per-frame peaks list.

    \item \textbf{ASIC-CCL pipeline.} For each frame, the
          \numproduct{11 x 11} patches reported by the on-chip CA
          stage undergo the same dark-subtraction, ring-masking, 
          intensity-thresholding and pseudo-Voigt-fitting pipeline as the software path.
\end{itemize}

The output of both pipelines is a per-frame peaks list in the standard
MIDAS format.

\subsection{Dataset}
\label{subsec:eval-dataset}

We evaluate the two pipelines on a FF-HEDM scan of an unloaded 304L austenitic stainless steel sample exposed to monochromatic, high-energy, synchrotron X-ray (71.68 keV) in a transmission geometry with a GE 41RT flat panel area detector (2048 $\times$ 2048 pixels with pixel size of 200 micron) placed at a distance of 803 mm from the sample. The sample was rotated about the vertical axis (normal to beam direction and aligned to loading axis) using an $\omega$ step size of 0.25 degrees. More details about the experimental setup are available in Park et al. \cite{park_repeatability_2021} and Zheng et al. \cite{zheng_rapid_2024}.

\subsection{Results: Grain-Level Equivalence}
\label{subsec:eval-grains}

\begin{table}[!htbp]
\centering
\caption{End-to-end equivalence of the two MIDAS reconstruction pipelines: Software-CCL and ASIC-CCL. ``$\Delta$'' is the ASIC-CCL value relative to the software-CCL value.}
\label{tab:midas-equivalence}
\begin{tabular}{l S[table-format=6.3]}
\hline
{Quantity}                                  & {Value}      \\
\hline
Peaks (Software-CCL)                         & 122296       \\
Peaks (ASIC-CCL)                                  & 121367       \\
$\Delta$ peaks (\%)                         & -0.76        \\
Per-frame peak Pearson $r$                  & 0.999      \\
\hline
Grains (Software-CCL)                        & 978          \\
Grains (ASIC-CCL)                                 & 980          \\
$\Delta$ grains (\%)                        & +0.20        \\
Mean grain radius, Software-CCL (\unit{\um}) & 75.13        \\
Mean grain radius, ASIC-CCL (\unit{\um})          & 75.21        \\
\hline
Mean strain RMS, Software-CCL (\unit{\micro\epsilon}) & 620.2 \\
Mean strain RMS, ASIC-CCL (\unit{\micro\epsilon})         & 620.0  \\
\hline
Fraction misorientation $<0.5^\circ$ (\%)   & 100.0        \\
Fraction centroid offset $<50$\,\unit{\um} (\%) & 100.0     \\

\hline
\end{tabular}
\end{table}

\begin{figure}[!htbp]
    \centering
    \includegraphics[width=0.75\textwidth]{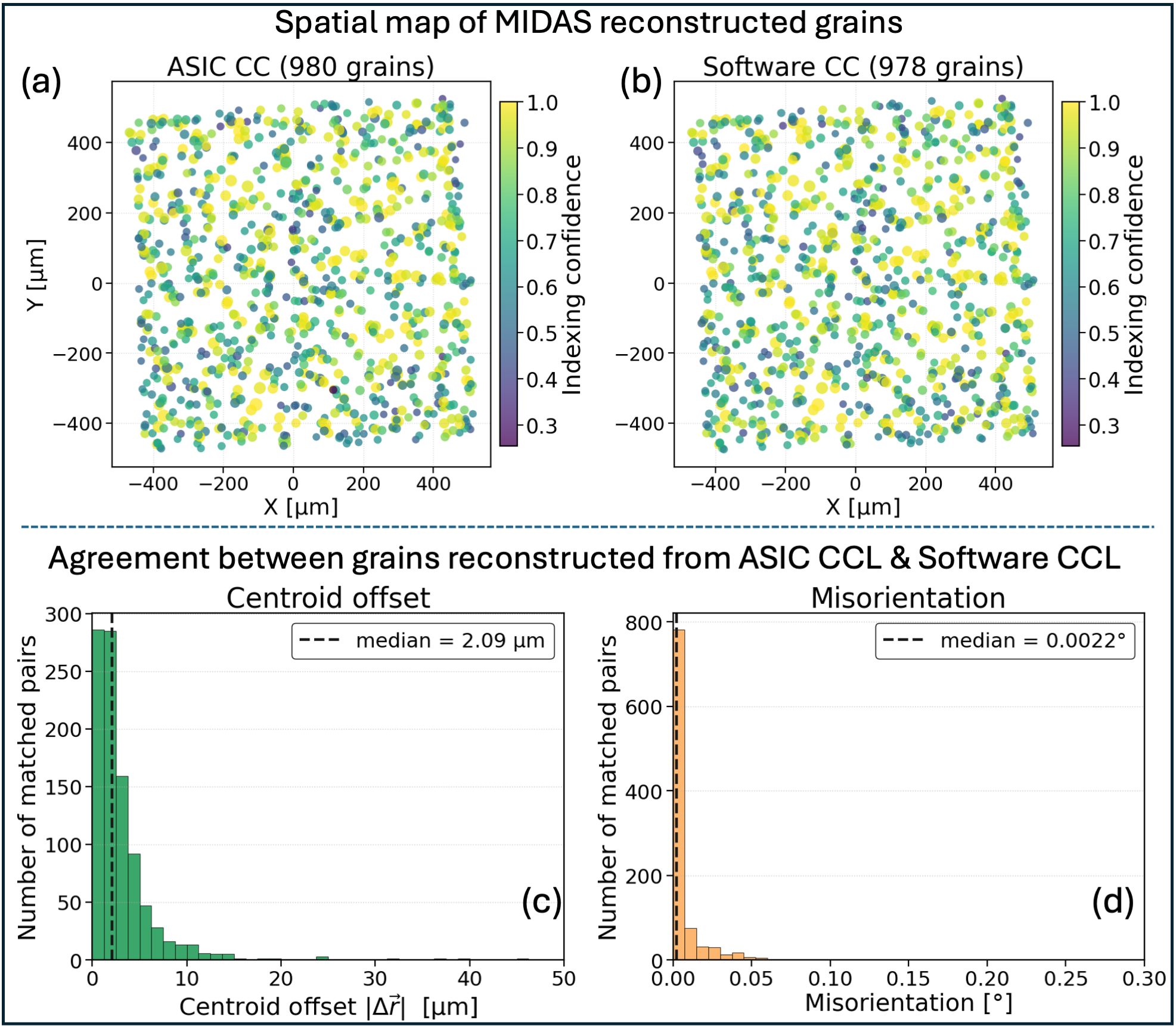} 
    \caption{MIDAS grain reconstruction: Spatial map of reconstructed grain from (a) ASIC-CCL and (b) Software-CCL pipelines. Distribution of (c) offer in grain centroid positions and (b) grain misorientation between the ASIC-CCL and Software-CCL grains.}
    \label{fig:midas}
\end{figure}

The indexed-grain populations agree to within $0.2\%$
(Table~\ref{tab:midas-equivalence}): \num{980} grains from the CA
pipeline versus \num{978} from the software-CC pipeline. The mean
grain radius (\qty{75.21}{\um} vs \qty{75.13}{\um}), mean indexing
confidence, and mean refined elastic-strain RMS
(\qty{620.0}{\micro\epsilon} vs \qty{620.2}{\micro\epsilon}) are
indistinguishable. The spatial maps of grain centroids in the
$(X,Y)$ plane (Fig.~\ref{fig:midas}) overlay one another for almost all the grains. The median centroid offset between the two
pipelines is \qty{1.85}{\um}---an order of magnitude below the
\qty{75.2}{\um} mean grain radius---and the median misorientation is
$0.0022^\circ$ (mean $0.0062^\circ$). $99.9\%$ of matched grains agree to within $0.1^\circ$ and $100\%$ agree to within $0.5^\circ$ misorientation. Similarly, $96.0\%$ agree to within \qty{10}{\um} in position and $100\%$ agree to within \qty{50}{\um} (Fig.~\ref{fig:midas}).

Since the downstream processing is identical for both MIDAS pipelines, the strong end-to-end equivalence is a direct consequence of the similar patches extracted by both the Software-CCL and ASIC-CCL pipelines, with differences in patch count aligning with the ASIC implementation filtering patches near the edge of the image that the Software-CCL would detect. The on-chip CA stage is therefore functionally equivalent to
software CC for downstream FF-HEDM reconstruction wherein the bandwidth saving from
transmitting compact \numproduct{11 x 11} instead of full
detector frames comes at no significant cost to reconstruction
quality.

\section{Conclusion}\label{sec:conclusion}
This work presents a hardware oriented approach to X-ray Bragg peak detection using a cellular automaton based method. By performing peak identification directly on the sensor silicon, the proposed system eliminates the need to transfer full detector frames to external software pipelines, significantly reducing communication bandwidth requirements. This is particularly important in high throughput imaging environments, where bandwidth is a critical limiting factor.

The software RTL simulation results demonstrate that the CA based approach achieves complete coverage of peaks identified by traditional connected components methods, while also detecting additional candidate regions due to its less restrictive filtering criteria. Although this increased sensitivity introduces some redundant or over sized detections, the system remains robust to variations in peak morphology, including those observed under higher pressure conditions.

In addition, the proposed architecture enables a fully streaming implementation with high throughput, making it well suited for integration with modern high speed X-ray detectors as well as enabling integration with patch-based machine learning models that can be deployed at edge \cite{braggNN, Dishant-SuperResolution}. End-to-end MIDAS reconstructions driven by the CA-extracted patches agree with a conventional software-CC reference to within 0.2$\%$ in grain count and 0.5$^\circ$ in orientation. The bandwidth saving therefore comes at no measurable cost to reconstruction quality.

Future work will focus on: (i) improving patch quality by incorporating additional filtering mechanisms to eliminate partial or overlapping peaks, (ii) on-chip ring-aware filtering to drop the candidate clusters that lie outside the rings of interest before they consume output bandwidth, (iii) adaptive patch sizing to accommodate the broader peaks observed under mechanical loading, (iv) integration of a discrete wavelet transform (DWT) background-subtraction stage upstream of the CA thresholding front-end, making a combined DWT–CA pipeline a natural extension of the present work~\cite{DWT} and (v) coupling the CA-extracted patches with distributed edge inference such as the HeteroViT architecture~\cite{HeteroViT}, which maps a single-layer Vision Transformer onto the ASIC–FPGA–GPU detector hierarchy for real-time keep/discard decisions. These improvements aim to further close the performance gap with software based methods while preserving the efficiency and scalability advantages of the hardware implementation. These improvements aim to further close the performance gap with software based methods while preserving the efficiency and scalability advantages of the hardware implementation.

\appendix

\acknowledgments

This work is primarily supported by the U.S. Department of Energy (DOE) Office of Science, Advanced Scientific Computing Research and Basic Energy Sciences, Advanced Scientific Computing Research for DOE User Facilities award X-ray \& Neutron Scientific Center for Optimization, Prediction, \& Experimentation (XSCOPE). Additionally support from the AUREIS project (part of Microelectronics Energy Efficiency Research Center for Advanced Technologies (MEERCAT)) and the Morpheus project, supported by DOE BES/SUF's Accelerator and Detector R\&D program. Work was supported by the U.S. DOE Office of Science-Basic Energy Sciences, under Contract No. DEAC02-06CH11357 at Argonne and Contract No. DE-AC02-76SF00515 at SLAC. Work performed at the Center for Nanoscale Materials and Advanced Photon Source, both U.S. Department of Energy Office of Science User Facilities (SUF), was supported by the U.S. DOE, Office of Basic Energy Sciences. Work at University of Chicago is supported by the Divisions of Chemistry (CHE) and Materials Research (DMR), National Science Foundation, under grant numbers NSF/CHE-1834750 and NSF/CHE-2335833. We gratefully acknowledge the computing resources provided and operated by the Joint Laboratory for System Evaluation (JLSE) at Argonne National Laboratory.

\bibliographystyle{JHEP}
\bibliography{references}

\end{document}